\documentclass[english]{article}
\usepackage[T1]{fontenc}
\usepackage[latin9]{inputenc}
\usepackage[letterpaper]{geometry}
\usepackage{amstext}
\usepackage{amssymb,amsfonts,amsmath}
\usepackage{epsfig}
\usepackage{amsthm}
\usepackage{amssymb}
\usepackage{amsmath}
\usepackage{graphics}
\usepackage{graphicx}
\usepackage{dcolumn}
\usepackage{bm}
\usepackage{subfigure}
\usepackage{xcolor}
\usepackage{color}
\usepackage[square,sort,comma,numbers]{natbib}

\makeatletter
\newcommand{\lyxaddress}[1]{
\par {\raggedright #1
\vspace{1.4em}
\noindent\par}
}

\makeatother

\usepackage{babel}

\begin{document}

%
%

\title{Evolution of cooperation driven by active information spreading}

\date{}

\author{Bin Wu$^{1}$, Hye Jin Park $^3$, Lingshan Wu$^2$, and Da Zhou$^{2,3}$}

\maketitle

\lyxaddress{$^1$School of Sciences, Beijing University of Posts and Telecommunications, Beijing 100876, PR China}
\lyxaddress{$^2$School of Mathematical Sciences, Xiamen University, Xiamen 361005, PR China}
\lyxaddress{$^3$Department of Evolutionary Theory, Max Planck Institute for Evolutionary
Biology, August-Thienemann-Str. 2, 24306 Pl\"{o}n, Germany}

\lyxaddress{
bin.wu@bupt.edu.cn (Bin Wu);\\
zhouda@xmu.edu.cn (Da Zhou).}


\begin{abstract}
Cooperators forgo their interest to benefit others.
Thus cooperation should not be favored by natural selection.
It challenges the evolutionists, since cooperation is widespread.
As one of the resolutions, information spreading has been revealed to play a key role in the emergence of cooperation.
Individuals, however, are typically assumed to be passive in the information spreading.
Here we assume that individuals are active to spread the information via self-recommendation.
Individuals with higher intensities of self-recommendation are likely to have more neighbors.
We find that i) eloquent cooperators are necessary to promote cooperation;
ii) individuals need to be open to the self-recommendation to enhance cooperation level;
iii) the cost-to-benefit ratio should be smaller than one minus the  ratio between  self-recommendation intensities of defector and cooperator,
which qualitatively measures the viscosity of the population.
Our results highlight the importance of active information spreading on cooperation.
\end{abstract}

%
%
%
%
%

\section{Introduction}
\label{}
Cooperators forgo their interest to benefit others.
Thus it cannot be favored by natural selection without additional mechanisms.
Cooperation, however, is ubiquitous ranging from genes to multicellularities in biology.
In addition, human society is based upon cooperation as well.
It has taken decades to fill the gap between evolutionary theory and widespread cooperation \cite{axelrod1981evolution,nowak2006five,sigmund2010calculus,nowak2011supercooperators}.
Evolutionary game theory takes natural selection into account as the driving force of evolution.
It provides a convenient paradigm to study the evolution of cooperation \cite{nowak2006evolutionary,hofbauer2003evolutionary}.
In particular, the Prisoner's Dilemma (PD) has been extensively adopted as a metaphor to study the emergence of cooperation \cite{axelrod1980effective,axelrod1980more,nowak1993strategy,cable1997prisoner}.
In the simplified PD game, a cooperator offers its opponent a benefit $b$ at a personal cost of $c$ ($b > c>0$), whereas a defector offers nothing.
As a result, it is best to defect irrespective of the co-player's decision.
In addition, it is shown that defection is the only evolutionary stable strategy (ESS) \cite{smith1982evolution,cressman2013stability} of the PD game with the replicator equation \cite{taylor1978evolutionary} which describes the dynamics of cooperation level in the well-mixed population.
Cooperation cannot be achieved even though mutual cooperation is the best for the group interest because all individuals try to maximize their own interest.
As thus, the PD game captures the conflict between group and individual interests so called \emph{social dilemma}.

To resolve this social dilemma, many mechanisms have been proposed, and one of the key factors is information.
Information plays an important role in the evolution of cooperation. The decision making processes are based on a variety of information, such as the historical behaviors and payoffs of partners \cite{nowak1998evolution}.
For example, individuals with imitation rule not only make use of the information of their own payoffs, but also take the opponents' payoffs into account when making decisions \cite{wu2010universality,traulsen2006stochastic,wu2015fitness}.
On the contrary, individuals with aspiration rule only make use of the information of their own payoffs to make decisions.
Due to this difference, a strategy favored by imitation rule can be disfavored by aspiration-based rule \cite{du2014aspiration,wu2018individualised}.
Thus, it is of importance what information individuals possess in decision making.
Once the information is available,
it owes to the process of information transmission and acquisition
to determine who obtains the true information.
This process is highly complex in the real world, and it is still unclear how the processes of information transmission and acquisition reshape the evolutionary dynamics.
To examine the effect of active information spreading on the emergence of cooperation,
we focus on self-recommendation, i.e., a self-promoting action to attract others.
On the one hand, commercial advertisement is of self-recommendation in human society \cite{presbrey2000history}.
Enterprises make advertisements on the mass media including TV, newspapers and internet to recommend their products to potential consumers.
In the animal world, courtship display is also of self-recommendation, that animals attract mates by showing off their beauty or strength \cite{hebets2004attention,martin2008object}.
For example, peacocks spread their beautiful tails to attract peahens \cite{beauchamp2014calling}.
Individuals self-recommending themselves are essentially active to deliver information and draw attention from others
in both animal worlds and human societies.
Thus self-recommendation is a widespread active way to spread information.
On the other hand, it can pave the way for individuals to figure out who the cooperator is via self-recommendation.
Surely it can be even more difficult to figure out who the cooperator is, if exaggerated or even false advertising is present.
For example,
the products or services can be not as good as stated in advertisements,
and consumers can be cheated via the self-recommendation.
Thus, it is of great importance how individuals respond to the self-recommendation.

To this end, we take the self-recommendation as the way of information transmission.
We assume that individuals can react to the self-recommendation or not as the way of information acquisition.
To be more precise,
we assume that both cooperators and defectors are able to recommend themselves when an individual decides to have a new neighbor.
Their capabilities of self-recommendation are characterized by self-recommendation intensities.
The larger the self-recommendation intensity is, the more eloquent the individual is,
and the more likely it is connected.
On the other hand, we introduce stubbornness to capture how convincible an individual is.
The individual with larger stubbornness value is less likely to accept the self-recommendation.
Our results show that cooperation is more likely to prevail,
provided that i) cooperators are more eloquent than defectors;
and that ii) individuals who are making decision to alter social partners are more convincible.
The results show that the self-recommendation reshapes the evolutionary fate of cooperation.

\section{Model}
We consider a structured population of $N$ individuals.
Initially the population is located on a regular network,
in which nodes represent individuals and links represent social ties between individuals.
We assume that the population size $N$ is much larger than the average degree $\langle d \rangle$ of the network ($N\gg \langle d \rangle $).
In other words, each individual's neighborhood has a limited size.
Each individual is either a cooperator ($C$, denoted by a unit vector $\textbf{s}=[1,0]^T$) or a defector ($D$, $\textbf{s}=[0,1]^T$).
The payoff matrix $\textbf{Q}$ is given by
\begin{equation}
\begin{array}{ccc}
 &\begin{array}{ll}
   C & \ \ \ \ \ D \end{array}\\
\begin{array}{c}C\\D\end{array}&\left(\begin{array}{cc}
$$b-c$$\  &$$-c$$\\
$$b$$\ \ &$$0$$\\
\end{array}\right),\end{array}
\label{PD}
\end{equation}
where $b>c>0$, and $c/b$ is the cost-to-benefit ratio.

At each time step, either the strategy of an individual or the
structure of the network is updated \cite{pacheco2006coevolution,wu2010evolution, perc2010coevolutionary,fu2008reputation,wu2011evolutionary,wu2016evolving}.
Let $\omega$ be the probability of strategy update, then $1-\omega$ corresponds
to the probability of network update.
The probability $\omega$ captures the relative time scales of the two processes.

\textbf{\emph{Strategy update}}. We adopt the Fermi updating rule \cite{traulsen2007pairwise,wu2010universality}. At each step of strategy update,
a focal individual $F$ is randomly selected from the population, and its accumulated payoff is calculated as
$
\Phi_F=\sum_{i\in \Delta_F} \textbf{s}_F^T \textbf{Q} \textbf{s}_i,
$
where $\Delta_F$ represents the neighborhood of individual $F$, and $\textbf{Q}$ is the payoff matrix.
The strategies of individuals $F$ and $i$ are represented by $\textbf{s}_F$ and $\textbf{s}_i$, respectively.
Then another individual $G$ is randomly selected among the neighborhood of $F$. The accumulated payoff of $G$ is given by
$\Phi_G=\sum_{i\in \Delta_G} \textbf{s}_G^T \textbf{Q} \textbf{s}_i$.
The focal individual $F$ compares its accumulated payoff $\Phi_F$ with $\Phi_G$ and switches to the strategy of $G$ with probability
$\left(1+\exp\left[-\beta \left( \Phi_G-\Phi_F \right) \right] \right)^{-1}$.
Non-negative $\beta$ controls the intensity of selection, which corresponds to an inverse temperature in statistical physics \cite{traulsen2007pairwise}.
Small $\beta$ implies weak selection.
In this case, individuals imitate other's strategies with probability approximately one-half,
even when the opponent gains much more than the focal individual.
In particular, zero selection intensity corresponds to the neutral drift \cite{kimura1983neutral}.
Large $\beta$ means strong selection.
In this case, individual $F$ is almost sure to adopt the strategy of individual $G$, provided individual $G$ gains even slightly more than individual $F$.
The infinite large selection intensity mirrors the perfect rationality in economics \cite{fudenburg1991game}.

\textbf{\emph{Network update}}.
At each step of network update, a link is randomly selected from the network and the link breaks off with probability $k$.
If it is broken, an individual between the two endpoints of the link is picked randomly as an active individual.
The active individual is to reform a new link, i.e. rewire to a new individual who is not in its current neighborhood.
On the one hand, we assume that all the individuals in the population are informed that the active individual is searching for a new partner.
And all the qualified potential neighbors, who are not in the active individual's neighborhood, recommend themselves to the active individual based on their intensities of recommendation.
For simplicity, we assume positive constants $R_C$ and $R_D$ to capture the intensities of self-recommendation for cooperators and defectors, respectively.
On the other hand, we also assume the stubbornness $p$ of an active individual.
With probability $p$, the active individual does not take account of others' self-recommendation \cite{burghardt2016competing}.
In this case, a new neighbor is randomly chosen regardless of self-recommendation.
Otherwise, the active individual does respond to the self-recommendation with probability $1-p$.
The active individual rewires to an individual with a probability proportional to its intensity of self-recommendation.
Consequently,  the active individual is likely to be attracted by eloquent individuals, those with large intensities of recommendation.

The network update captures the process of  information transmission and its acquisition.
For the information transmission, self-recommendation is adopted.
Individuals compete for social ties, as the Moran process does \cite{moran1962statistical}.
The intensity of the self-recommendation mirrors the fitness of an individual, and links mirror the off-springs.
Individuals with higher self-recommendation intensities have larger likelihood to become the new neighbor of the active individual,
as individuals with larger fitness are more likely to produce offsprings.
Furthermore, the total number of links  keeps constant over time,
as the population size keeps invariant in the Moran process.
For the information acquisition, we assume a react-or-non-react model.
The probability to react to the self-recommendation mirrors the selection strength in the Moran process:
Individuals with high probabilities of reacting to the self-recommendation is likely to connect with individuals with strong intensities of self-recommendation,
as individuals with large fitness are very likely to reproduce only if the selection intensity is strong in the Moran process.
Therefore, the network update is also an evolutionary process.

\section{Analysis}
In this section, we make use of mean-field analysis to show how cooperative behavior is reshaped by the self-recommendation,
which drives the topology to evolve.
Noteworthy, both the strategy and network structure evolve.
It gives rise to a coupled dynamics,
which is typically challenging to solve \cite{gross2008adaptive}.
We overcome this  by assuming $\omega\ll 1$, i.e. the network update is much more frequent than the strategy update (see \ref{appB}).
Thus, the network structure keeps evolving, and reaches its stationary regime before individuals update their strategies.
In this case, the linking dynamics is captured by a Markov chain.
The resulting stationary distribution of the Markov chain $\vec{\pi}_0$ quantitatively indicates the fraction of $CC$, $CD$ and $DD$ links of the network, respectively.
The stationary distribution is given by
\begin{equation}
\vec{\pi}_0=(\pi_{01},\pi_{02},\pi_{03})=\chi^{-1}(\gamma_3\gamma_4,\gamma_1\gamma_4,\gamma_1\gamma_2),
\label{stationary}
\end{equation}
where
\begin{align}
\gamma_1&=pkx_D(x_CR_C+x_DR_D)+(1-p)kx_DR_D,\nonumber\\
\gamma_2&=\frac{1}{2}kx_Dp(x_CR_C+x_DR_D)+\frac{1}{2}(1-p)kx_DR_D,\nonumber\\
\gamma_3&=\frac{1}{2}kx_Cp(x_CR_C+x_DR_D)+\frac{1}{2}(1-p)kx_CR_C,\nonumber\\
\gamma_4&=pkx_C(x_CR_C+x_DR_D)+(1-p)kx_CR_C,\nonumber\\
\label{stationary1}
\end{align}
and $\chi=(\gamma_3\gamma_4+\gamma_1\gamma_4+\gamma_1\gamma_2)$
is the normalization factor (see \ref{appB}).

If the active individuals are stubborn, i.e. $p=1$,
the stationary distribution becomes $(x_C^2,2x_Cx_D,x_D^2)$.
This is  the same as that in the well-mixed population.
If the active individuals are not stubborn at all, i.e., $p=0$,
the stationary distribution $\vec{\pi}_0$ is given by
$(\alpha_C^2, 2\alpha_C\alpha_D, \alpha_D^2)$
with $\alpha_s=R_sx_s(R_Cx_C+R_Dx_D)^{-1}$, $s\in\{C,D\}$.
It implies that the self-recommendation reshapes the population structure.
It acts as if the well-mixed population with a rescaled frequency of cooperators,
i.e. from $x_C$ to $\alpha_C$.
To be precise, it implies that i) $\pi_{01}$ is a monotonically increasing function of $R_C/R_D$, that is, the fraction of $CC$ links increases with $R_C/R_D$; ii) $\pi_{03}$ is a monotonically decreasing function of $R_C/R_D$, and the fraction of $DD$ links decreases with $R_C/R_D$; iii) $\pi_{02}$, i.e., the fraction of $CD$ links increases at first and then decrease with $R_C/R_D$.
Note that there are few $CD$ links, provided $R_C/R_D$ is large enough.

The stationary regime of the network structure facilitates us to estimate the average accumulated payoff of both cooperators and defectors.
If we additionally assume that the population size is sufficiently large,
we find that the fraction of cooperation is approximately captured by
\begin{equation}
\dot{x_C}=x_C(1-x_C)\left[(R_C-R_D)((1-p)b-c)x_C-R_Dc\right],
\label{model}
\end{equation}
which can be found in \ref{appA}.

Noteworthy, Eq.~\eqref{model} is equivalent to the replicator dynamics $\dot{x}_C=x_C(1-x_C)(\tilde{f}_C-\tilde{f}_D)$ \cite{taylor1978evolutionary}, in which $\tilde{f}_C$ and $\tilde{f}_D$ are determined by the transformed payoff matrix $\tilde{\textbf{Q}}$
\begin{align}
\begin{array}{ccc}
 &\begin{array}{ll}
   C & \ \ \ \ \ \ \ \ \  \ \ \ \ \ \ \ \ \ \ \ \ D \end{array}\\
\begin{array}{c}C\\D\end{array}&\left(\begin{array}{cc}
$$R_C(1-p)(b-c)$$\  &$$-R_Dc$$\\
$$R_D(1-p)b+R_Cpc$$\ \ &$$0$$\\
\end{array}\right).
\end{array}
\label{transform}
\end{align}
In other words, self-recommendation essentially changes the interaction between cooperators and defectors \cite{nowak2006five}.
The transformed payoff matrix Eq.~\eqref{transform} captures the interaction between cooperators and defectors shifted by self-recommendation.

\section{Results}
In this section, we make use of the replicator equation with the transformed game Eq.~\eqref{transform} to investigate how self-recommendation intensities ($R_C$ and $R_D$) and stubbornness ($p$) affect the evolutionary dynamics of cooperation.
Besides, we also investigate the robustness of the results with respect to the parameters which are absent in the transformed game Eq.~\eqref{transform} via simulations.

\subsection{More eloquent cooperators and less stubborn individuals promote cooperation}
\begin{figure}
\begin{center}
\includegraphics[width=0.8\textwidth]{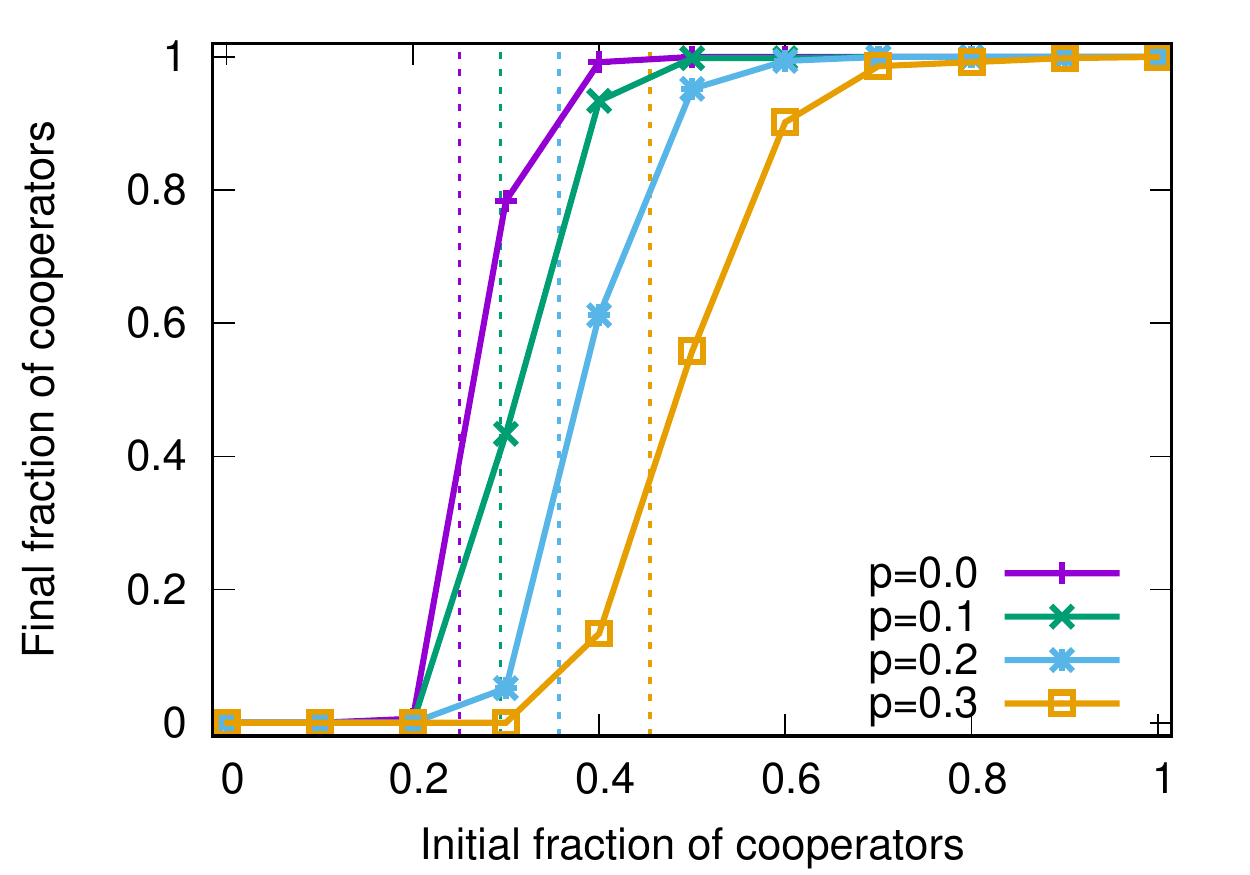}
\caption{The final fraction of cooperators as a function of initial fraction of cooperators.
The symbols indicate the simulation results and the dashed lines represent the internal unstable fixed point $x^*_C$ given by Eq.~\eqref{x_C}.
The simulation results are in agreement with our theoretical predictions.
If the initial value is small than the critical fraction of cooperators $x^*_C$, defection dominates the population.
Otherwise, cooperation is taking over the population.
This mirrors a coordination game with cooperation a stable Nash equilibrium.
Each data point in this figure is averaged over $500$ independent runs.  In each run, we set a transient time of $10^6$ generations.
(Parameters: Prisoner's Dilemma with $b=3$ and $c=1$. The intensities of self-recommendation of cooperators and defectors, $R_C=3$ and $R_D=1$ respectively. Population size $N=1000$, average degree $\langle d\rangle=20$, probability of a strategy update $\omega=10^{-3}$, link-breaking probability $k=1$, and selection intensity $\beta=10$.)}
\label{Fig1}
\end{center}
\end{figure}

We resort to the transformed matrix Eq.~\eqref{transform} to shed light on the evolutionary fate of cooperation.
If the opponent is a defector,
then the effective payoff of a defector, i.e., $0$, must be greater than that of a cooperator, i.e., $-R_Dc<0$.
If the opponent is a cooperator,
a cooperator is better off than a defector if and only if $R_C(1-p)(b-c)>R_D(1-p)b+R_Cpc$ holds.
In other words, cooperation is a strict Nash Equilibrium of the transformed matrix,
provided that both
\begin{equation}
R_C>R_D~~~~\textrm{and}~~~~p<1-\frac{c}{b}\frac{1}{1-\tfrac{R_D}{R_C}}
\label{pvalue}
\end{equation}
hold, which is equivalent to the inequality $R_C(1-p)(b-c)>R_D(1-p)b+R_Cpc$.
In this case, the transformed matrix is a coordination game,
in which individuals are better off to do what others do.
Both $x_C^*=1$ and $x_C^*=0$ are stable fixed points for Eq.~\eqref{model}, separated by the unstable internal fixed point
\begin{equation}
x_C^*=\frac{cR_D}{(R_C-R_D)((1-p)b-c)}.
\label{x_C}
\end{equation}
Therefore, if Eq.~\eqref{pvalue} is fulfilled,
cooperation dominates the population provided that the initial fraction of cooperators exceeds the critical value $x^*_C$ by Eq.~\eqref{x_C}.
Otherwise, defection takes over the population.
The critical value $x_C^*$ via simulation is in good agreement with the theoretical prediction (Fig~\ref{Fig1}).

Based on Eq.~\eqref{pvalue},  two conditions are required to promote cooperation:
i) the cooperators should be better than defectors at self-recommendation, i.e., $R_C>R_D$;
ii) individuals should be less stubborn and more open to self-recommendation, i.e., $p<1-\frac{c}{b}\frac{1}{1-\tfrac{R_D}{R_C}}$.

For i), it indicates that cooperation emerges only if cooperators are more eloquent than defectors.
On the one hand, if $R_C>R_D$  cooperators are more active and are more likely to be selected as a new neighbor.
This would make cooperators clustered together.
The clustered cooperators interact more often with each other and gain higher payoffs.
Thus they would outperform defector neighbors in payoff,
and eventually take over the population \cite{pacheco2006coevolution}.
Noteworthy, eloquent cooperators are taking risks because
they would attach to defectors from time to time.
However, the resulting clustered cooperators expand so quickly that the risk is under control.
On the other hand, let us consider the situation that $R_C>R_D$ does not hold. We focus on a special case $R_C= R_D$, i.e. cooperators and defectors are equally eloquent.
In this case, all the potential new neighbors of the active individual have  the same likelihood to be selected.
In other words, there is no preferential attachment in the linking dynamics.
The stationary population regime ends up with $\pi_0=(x_C^2,2x_Cx_D,x_D^2)$,
which is the same as that in the well-mixed population.
The resulting replicator equation is $\dot{x_C}=-cx_C(1-x_C)$,
which is exactly the replicator dynamics of the original PD game.
As a result, defection dominates the population regardless of the  stubbornness $p$.
Therefore $R_C>R_D$ is necessary for the emergence of  cooperation.
\begin{figure}
\begin{center}
\includegraphics[width=1\textwidth]{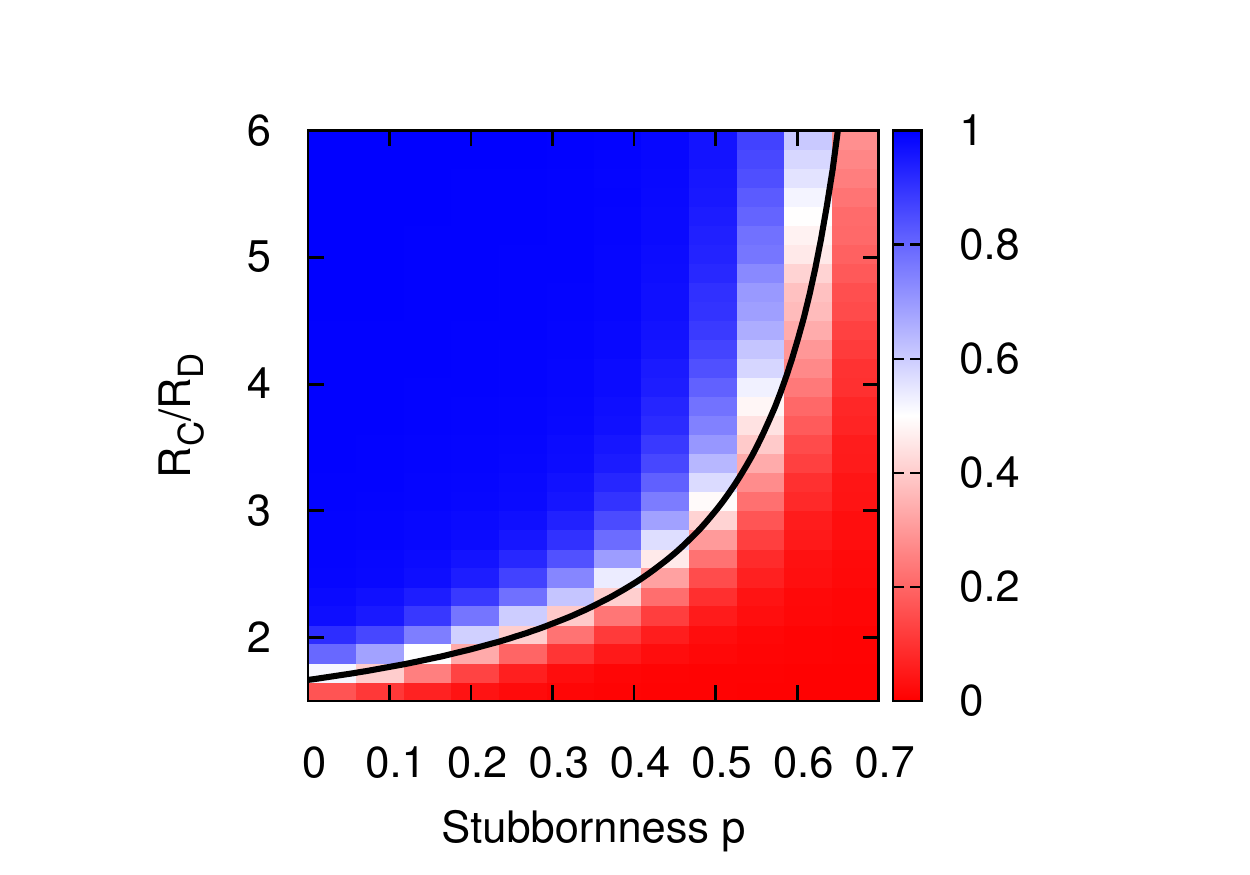}
\caption{The final fraction of cooperators as a function of $R_C/R_D$ and stubbornness $p$.
It is shown that cooperation prevails if the cooperators are more eloquent, i.e., $R_C/R_D$ is sufficiently large;
and if individuals are open to accept the self-recommendation, i.e., $p$ is small.
The simulation, i.e., the heat map, is in agreement with our theoretical prediction, i.e., the black line determined by Eq~\eqref{x_C}. (Parameters: Prisoner's Dilemma with $b=3$ and $c=1$, $N=1000$, $\langle d\rangle=20$, $\omega=10^{-3}$, $k=1$, and $\beta=10$. The initial fraction of cooperators is $0.5$.)}
\label{Fig2}
\end{center}
\end{figure}

For ii), individuals should be open and less stubborn, i.e., $p<1-\frac{c}{b}\frac{1}{1-\tfrac{R_D}{R_C}}$ (see Fig.~\ref{Fig2}).
To illustrate this,
let us consider two extreme cases: $p=1$ and $p=0$.
When $p=1$, i.e. the active individual is so stubborn that it does not react to the self-recommendation by anybody.
In this case, the active individual randomly chooses new neighbors.
The self-recommendation does not work.
The resulting stationary population regime is the same as that of the well-mixed population.
Thus the approximated replicator equation is  $\dot{x_C}=-cx_C(1-x_C)$,
as that in the well-mixed population.
Therefore, defection dominates the population.
\begin{figure}
\begin{center}
\includegraphics[width=1\textwidth]{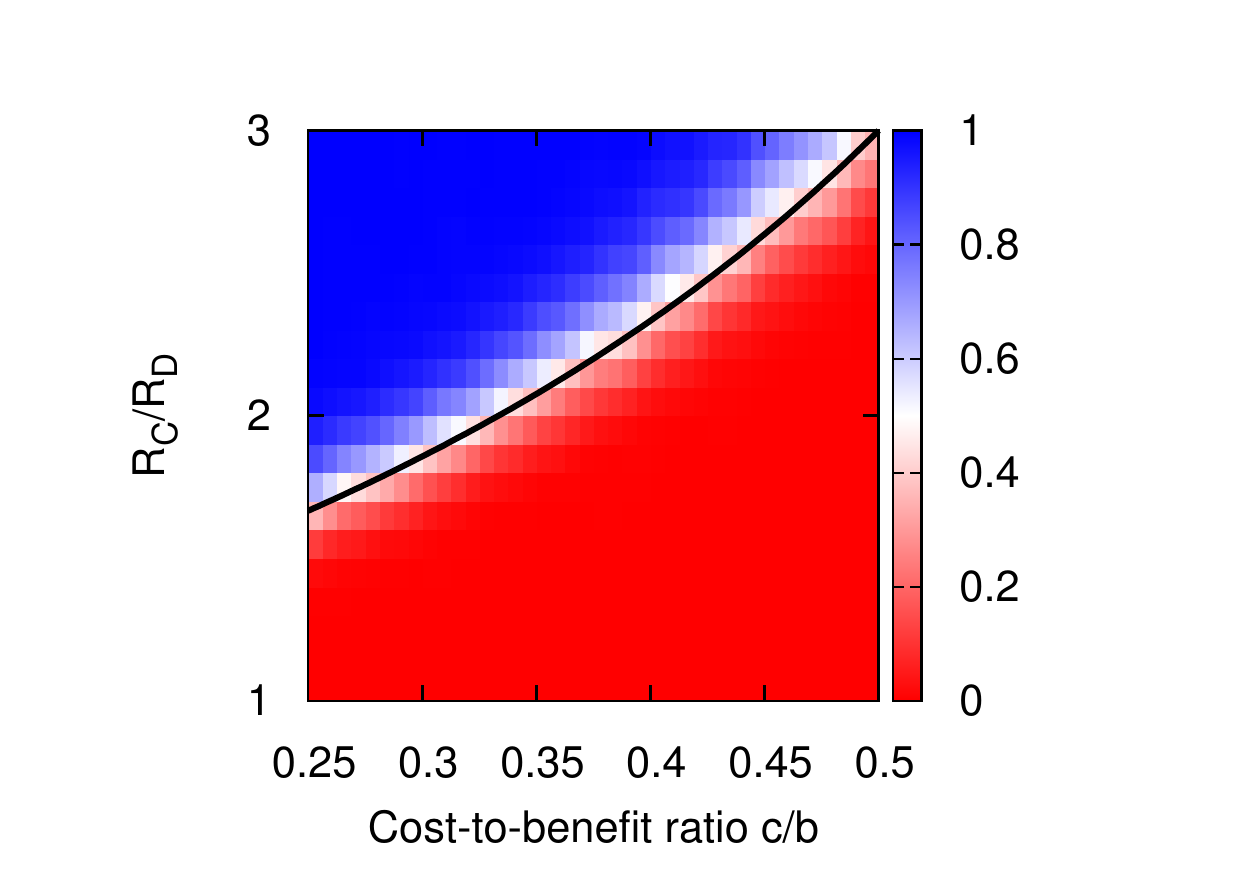}
\caption{The final fraction of cooperators as a function of $R_C/R_D$ and cost-to-benefit ratio $c/b$.
The heat map represents the simulation result and the black line indicates the theoretical boundary predicted by $x^*=cR_D/[(b-c)(R_C-R_D)]=0.5$. We can see that the simulation is in agreement with our theoretical prediction.
It is shown that small cost-to-benefit ratio facilitates the cooperation.
(Parameters: $N=1000$, $\langle d\rangle=20$, $\omega=10^{-3}$, $k=1$, and $\beta=10$. The initial fraction of cooperators is $0.5$ and the stubbornness $p=0$.)}
\label{Fig3}
\end{center}
\end{figure}
Let us resort to the other extreme case $p=0$, i.e. individuals are not stubborn at all, and take into account of the self-recommendation.
The transformed payoff matrix becomes
\begin{align}
\left(\begin{array}{cc}R_C(b-c)&-R_Dc\\R_Db&0\end{array}\right).
\label{trans2}
\end{align}
The transformed payoff matrix Eq.~\eqref{trans2} shows that
a) when a defector meets a defector, the effective payoff of each defector is still $0$ as in the original PD game Eq.~\eqref{PD};
b) when a defector meets a cooperator, both individuals obtains $R_D$ times  the payoff of the original PD game;
c) when a cooperator meets a cooperator, each cooperator gets $R_C$ times the payoff of the original PD game.
Noteworthy, the group interest of two cooperators $2R_C(b-c)$
outperforms that of a cooperator and a defector $R_D(b-c)$, provided $2R_C>R_D$.
Thus cooperation could still be a social optimum as in the original PD game Eq.~\eqref{PD}.
For the emergence of cooperation, it is essential to compare $R_C(b-c)$ and $R_Db$.
In the original PD game, $b-c$ is less than $b$, whereas $R_C(b-c)$ can be larger than $R_Db$ as long as $R_C$ is sufficiently larger than $R_D$.
To be precise, if
\begin{equation}
\frac{c}{b}<1-\frac{R_D}{R_C},
\label{simplerule}
\end{equation}
cooperators take over the whole population as long as the initial fraction of cooperators exceed $x^*_C=cR_D/[(b-c)(R_C-R_D)]$.
The cost-to-benefit ratio $c/b$ is smaller than $1$, thus  $R_C>R_D$ is necessary if Eq.~\eqref{simplerule} holds.
Or cooperation  prevails only if cooperators are much more eloquent than defectors (see Fig.~\ref{Fig3}).
For $p$ between zero and one,
the interaction between cooperators and defectors can be captured by Eq.~\eqref{transform}.
As $p$ increases, the effective game moves from Eq.~\eqref{PD},
in which cooperation is not a Nash equilibrium,
to Eq.~\eqref{trans2},
in which cooperation becomes an ESS.

\subsection{Robustness of theoretical predictions}

All the above results are based on the replicator equation of the transformed matrix Eq.~\eqref{transform}.
Eq. \eqref{transform} is determined by the cost-to-benefit ratio, the intensities of self-recommendation and the probability to react
to the self-recommendation.
However, the population size $N$, average degree $\langle d \rangle$, as well as the frequency of the strategy updates $\omega$ are absent in the transformed matrix.
Here we investigate the robustness of our theoretical predictions with respect to these parameters via simulations.

\begin{figure}
\begin{center}
\includegraphics[width=1\textwidth]{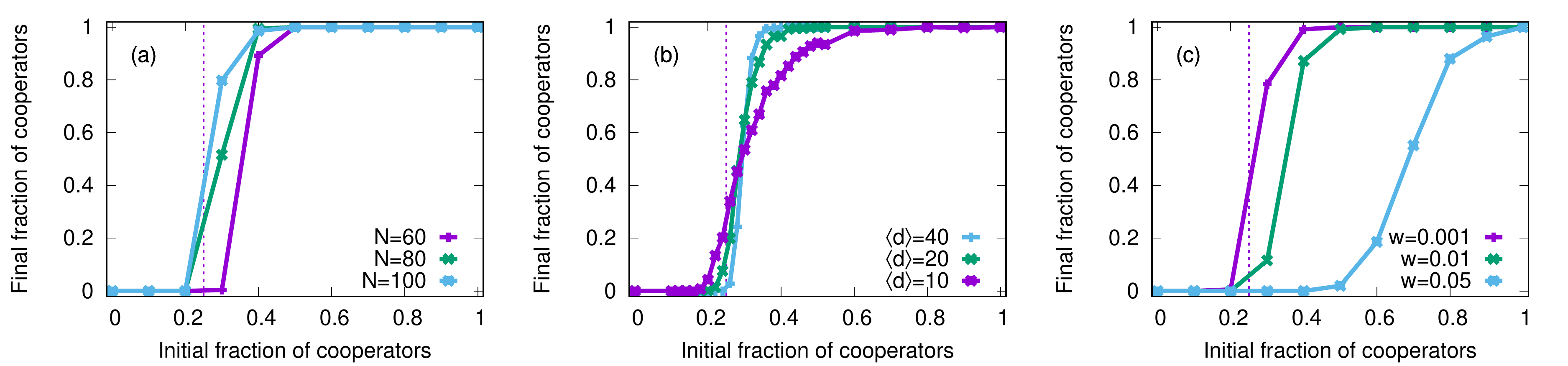}
\caption{Illustration of the robustness of our theoretical result.
(a) The left panel shows the simulation results for different population sizes. The critical transition from all defection to all cooperation is shifted to the right with the decrease of $N$.
(b) The middle panel shows the simulation results for different average degrees. The case with smaller average degree shows more fluctuations than the one with larger average degree.
(c) The right panel shows that the simulation results are quite sensitive to the frequency of the strategy updates $\omega$. The critical transition is shifted to the right with the increase of $\omega$.
The default parameters are the same as in Fig.~\ref{Fig1}.}
\label{Fig4}
\end{center}
\end{figure}

Firstly, we find that the larger the
population size is, the better the agreement is shown between
simulation results and theoretical prediction (Fig~\ref{Fig4}(a)). Noteworthy,
our theoretical prediction is based on the mean-field analysis
assuming that the population size is large enough.
Therefore, it is not surprised to see the disagreement when
the population size is small. Note that the critical value is
shifted to the right with the decrease of $N$ (Fig~\ref{Fig4}(a)).
Actually, given the average degree ( $\langle d \rangle=20$ in Fig~\ref{Fig4}(a)), smaller population size corresponds to denser network structure,
which could suppress the assortment of cooperators during the network update.

Secondly, we investigate the average degree $\langle d \rangle$ (Fig~\ref{Fig4}(b)). We can see that
the smaller $\langle d \rangle$ is, the more fluctuated the simulation result is.
In other words, the transition region from all defection to all cooperation becomes less sharp
as $\langle d \rangle$ decreases.
This phenomenon seems a little bit counterintuitive because our theoretical approximation is based on the assumption
that the whole population size is much larger than the average degree of the network. It means that, given the same population size,
the case with smaller average degree should better agree with the theoretical prediction. However, it should be noted that
smaller average degree increases the stochastic of the linking dynamics, because smaller $\langle d \rangle$ results in less
number of total links ($L=N\langle d \rangle/2$) in the system. In this way, properly increasing the value of $\langle d \rangle$ can reduce the fluctuations
around the expected numbers of $CC$, $CD$ and $DD$ links in the stationary regime, making the simulation results accord with
our theoretical prediction better.

Finally, let us discuss the parameter $\omega$, i.e. the frequency of the strategy updates (Fig~\ref{Fig4}(c)). We find that the simulation results
are sensitive to $\omega$. In particular, the critical transition from all defectors to all cooperators increases with the increase of $\omega$. Noteworthy, our method assumes that the network update is much faster than the strategy update ($\omega \ll 1$),
it thus makes sense that the simulation with smaller $\omega$ shows better agreement with our theoretical prediction. As
$\omega$ becomes larger, the strategy will update more frequently, and
it is not long enough for the network to reach the stationary regime.
Therefore, the population structure reshaped by the self-recommendation
would be of less benefit for cooperators to form clusters, and then
would inhibit the emergence of cooperation.

\section{Conclusions and Discussions}

Active information spreading is ubiquitous,
ranging from commercial advertisement \cite{presbrey2000history} to election campaign \cite{huckfeldt1995citizens}.
Those with high intensities of self-recommendation are likely to convince others.
It is similar to reputation,
where cooperators are typically assumed to have a high score of reputation \cite{nowak1998evolution,fu2008reputation,chen2009cooperation,gallo2015effects,ohtsuki2015reputation,park2017role,gross2019rise}.
Both self-recommendation and reputation play their role in the communication. Both cannot work unless the observers recognize.
However, they remarkably differ from each other.
On the one hand,
individuals take the initiative to attract others when they recommend themselves.
Those who self-recommend are active.
Reputation is at work in a passive way.
Individuals with high score of reputation could be too shy to be recognized.
On the other hand,
the reputation is driven by a social norm, and it is an opinion of a population.
The self-recommendation is of personal ability which is up to the focal individual.

We find that eloquent cooperators are necessary to promote cooperation.
The intuition is not straightforward, because an eloquent cooperator does not imply that cooperation is at an advantage:
If the active individual is a cooperator,
an eloquent cooperator with high $R_C$ would be more likely to  become the active cooperator's potential new neighbor.
Once the new link is established, it benefits both the active cooperator and the eloquent cooperator.
If the active individual is a defector, however,
an eloquent cooperator once again would be more likely to connect with the active defector.
This new link would only benefit the active defector rather than the cooperator.
In fact, the eloquent cooperators can form clusters, although they occasionally attach to defectors.
The clustered cooperators gain much more than their defector neighbors.
Thus the occasional attached defector would open an avenue for the cooperator cluster to spread.
Eventually cooperation takes over the whole population \cite{taylor2006evolutionary}.
Therefore, our results echo the so-called network reciprocity that cooperators can prevail by forming cooperative clusters \cite{nowak2006five}.

In addition, we find that the observers should be not too stubborn,
and they should be open to the self-recommendation.
In particular, if the individual are $100\%$ open to the self-recommendation,
cooperation becomes the stable Nash equilibrium,
provided $\frac{c}{b}<1-\frac{R_D}{R_C}$.
The larger $R_C$ is, the larger $1-\frac{R_D}{R_C}$ is, the more likely it is to form a cooperative cluster.
In other words, $1-\frac{R_D}{R_C}$ represent the social viscosity,
which is similar to \cite{ohtsuki2006simple,pacheco2006coevolution,wu2010evolution}.

To sum up,
our results suggest that cooperators should be encouraged to be more active in communicating with others,
otherwise the defectors would mislead and drive the population to the tragedy of the commons \cite{hardin1968tragedy}.


\section*{Acknowledgements}

BW acknowledges the sponsorship by the NSFC (Grants No.61603049, No.61751301). DZ is supported by the China Scholarship Council (No. 201806315038) and the Fundamental Research Funding for the Central Universities in China (No. 20720180005). Both HJP and DZ acknowledge the support from the Max Planck Institute for Evolutionary Biology. All the four authors acknowledge the
comments from the Department of Evolutionary Theory, Max Planck Institute for Evolutionary Biology.
\appendix

\section{Markov linking dynamics and its stationary distribution}
\label{appB}
For the linking dynamics,
there are three different types of social ties: cooperator-cooperator ($CC$), cooperator-defector ($CD$) and defector-defector ($DD$) links.
Based on the presented rewiring rule with self-recommendation, the network updating process can be modeled as a discrete-time Markov chain in the state space of $\{CC, CD, DD\}$ \cite{gardiner1985handbook,durrett2010probability}.
Let us denote $x_C$ and $x_D$ be the fraction of cooperators and defectors in the population.

We take the transition from $CD$ to $DD$ as an example.
This transition happens only when a $CD$ link is broken off (with probability $k$) and the $D$ individual is selected as the active individual (with probability $1/2$), and then rewire to another defector (with probability $x_D$ if $D$ does not respond or with probability $\alpha_D$ if it does respond). As a result, the transition probability is either $k\frac{1}{2}x_D$ or $k\frac{1}{2}\alpha_D$, depending on whether or not the active individual responds to the self-recommendation by others. According to the Law of Total Probability, the transition probability from $CD$ to $DD$ is given by $pk\frac{1}{2}x_D+(1-p) k\frac{1}{2}\alpha_D$.
All the rest entries of the transition matrix $\textbf{M}_0$ are obtained in the same argument.

We thus end up with the transition matrix
\begin{equation}
\textbf{M}_0=p\textbf{M}_1+(1-p)\textbf{M}_2,
\label{matrix}
\end{equation}
where
\begin{equation}
\textbf{M}_1=\begin{array}{cccc}
 &\begin{array}{lll}
   CC & \ \ \ \ \ \ \ \ \  \ CD & \  \  \ \ \ \
\ \ \ \ DD \end{array}\\
\begin{array}{c}CC\\CD\\DD\end{array}&\left(\begin{array}{ccc}
$$1-kx_D$$\  &$$kx_D$$&\  $$0$$\\
$$kx_C/2$$\ \ &$$1-(k/2)$$&\ \ $$kx_D/2$$\\
$$0$$\ \ &$$kx_C$$&\ \ $$1-kx_C$$
\end{array}\right)\end{array},
\end{equation}
and
\begin{equation}
\textbf{M}_2=\begin{array}{cccc}
 &\begin{array}{lll}
   CC & \ \ \ \ \ \ \ \ \  \ CD & \  \  \ \ \ \
\ \ \ \ DD \end{array}\\
\begin{array}{c}CC\\CD\\DD\end{array}&\left(\begin{array}{ccc}
$$1-k\alpha_D$$\  &$$k\alpha_D$$&\  $$0$$\\
$$k\alpha_C/2$$\ \ &$$1-(k/2)$$&\ \ $$k\alpha_D/2$$\\
$$0$$\ \ &$$k\alpha_C$$&\ \ $$1-k\alpha_C$$
\end{array}\right)\end{array}
\end{equation}
with $\alpha_C=x_CR_C/(x_CR_C+x_DR_D)$ and $\alpha_D=x_DR_D/(x_CR_C+x_DR_D)$.

Actually, $\textbf{M}_1$ and $\textbf{M}_2$ are the transition probability matrices conditional on the response and non-response cases respectively. $\textbf{M}_0$ is the convex combination of them due to Law of Total Probability.
The resulting Markov chain is aperiodic and irreducible, provided $x_Cx_D\neq 0$.
And there is a unique stationary distribution.
By solving the linear equation $\vec{\pi}_0 \textbf{M}_0=\vec{\pi}_0$, we obtain the stationary distribution $\vec{\pi}_0$ which is given by
\begin{equation}
\vec{\pi}_0=(\pi_{01},\pi_{02},\pi_{03})=\chi^{-1}(\gamma_3\gamma_4,\gamma_1\gamma_4,\gamma_1\gamma_2),
\label{stationary}
\end{equation}
where
\begin{align}
\gamma_1&=pkx_D(x_CR_C+x_DR_D)+(1-p)kx_DR_D,\nonumber\\
\gamma_2&=\frac{1}{2}kx_Dp(x_CR_C+x_DR_D)+\frac{1}{2}(1-p)kx_DR_D,\nonumber\\
\gamma_3&=\frac{1}{2}kx_Cp(x_CR_C+x_DR_D)+\frac{1}{2}(1-p)kx_CR_C,\nonumber\\
\gamma_4&=pkx_C(x_CR_C+x_DR_D)+(1-p)kx_CR_C,\nonumber\\
\label{stationary1}
\end{align}
and $\chi=(\gamma_3\gamma_4+\gamma_1\gamma_4+\gamma_1\gamma_2)$
is the normalization factor.

$\vec{\pi}_0$ characterizes the relative frequencies of different types of links when the network structure in the stationary regime.

\section{Replicator-like equation of the cooperation dynamics}
\label{appA}

Normally it is quite challenging to analyze the entangled dynamics
of strategy update and social relationship adjustment \cite{gross2008adaptive}.
Here we overcome this challenge by assuming that the network update is much faster than
the strategy update ($\omega \ll 1$).
In this case, the network structure reaches its stationary regime before a strategy update happens. This assumption facilitates us to obtain more tractable model approximation. In light of this, the idea of time scale separation has been frequently used in a variety of complex dynamics \cite{pacheco2006coevolution,wu2010evolution,wu2011evolutionary,
wu2016evolving,pastor2015epidemic,wu2016control,schwarzkopf2010epidemic,
guerra2010annealed}.

Note that there are three types of social ties. Let $N_{CC}$, $N_{CD}$ and $N_{DD}$ be the numbers of $CC$, $CD$ and $DD$ links. In the stationary regime we have
\begin{align}
N_{CC}=L \pi_{01}, \nonumber\\
N_{CD}=L \pi_{02}, \nonumber\\
N_{DD}=L \pi_{03}. \nonumber\\
\end{align}
$L=N\langle d \rangle/2$ is the total number of links, which remains unchanged during the network updates. $\vec{\pi}_0=(\pi_{01},\pi_{02},\pi_{03})$ represents the stationary distribution
given by Eq.~\eqref{stationary}. We then calculate the average
payoffs of cooperators and defectors respectively as follows
\begin{equation}\begin{array}{rcl}
f_C&=&(2(b-c)N_{CC}-cN_{CD})/Nx_C
\\&=& (2(b-c)\pi_{01}-c\pi_{02})L/Nx_C
\end{array}
\label{fc}
\end{equation}
and
\begin{equation}\begin{array}{rcl}f_D&=&bN_{CD}/Nx_D\\&=& bL\pi_{02}/Nx_D\end{array}.
\label{fd}
\end{equation}
When the population size $N$ is sufficiently large, the model with Fermi updating rule
can be captured by the following equation \cite{traulsen2006stochastic,wu2010evolution}
\begin{equation}\begin{array}{rcl}
\dot{x_C}
&=&x_C(1-x_C)\tanh(\beta(f_C-f_D)/2).
\end{array}
\label{meanfield}
\end{equation}
Let
\begin{equation}
G_0(x_C)=(f_C-f_D)^{-1}\tanh(\beta(f_C-f_D)/2),
\end{equation}
Eq.~\eqref{meanfield} can be rewritten as
\begin{equation}
\dot{x_C}=G_0(x_C)x_C(1-x_C)(f_C-f_D).
\label{rep}
\end{equation}
Note that $G_0(x_C)$ is positive, Eqs. \eqref{rep} and \eqref{meanfield}
have the same fixed points and stability properties. In other words, they are equivalent to each other in terms of evolutionary stability. Therefore, Eq.~\eqref{rep} captures the evolution of cooperation in our model.
Substituting Eqs. \eqref{fc} \eqref{fd} \eqref{stationary} and  \eqref{stationary1} into Eq.~\eqref{rep} leads to:
\begin{equation}
\dot{x_C}=Q(x_C) x_C(1-x_C)\left[(R_C-R_D)((1-p)b-c)x_C-R_Dc\right],
\label{last}
\end{equation}
where
\begin{equation}
Q(x_C)=\frac{Lk^2}{N\chi}(p(x_CR_C+x_DR_D)+(1-p)R_C)
\end{equation}
is a positive rescaling factor. Therefore, we simplify \eqref{last} as
\begin{equation}
\dot{x_C}=x_C(1-x_C)\left[(R_C-R_D)((1-p)b-c)x_C-R_Dc\right]
\label{last2}
\end{equation}
without changing its evolutionary stability.

%
%
%


\end{document}